\documentclass[]{article}

\usepackage[T1]{fontenc}		

\usepackage{amsmath,amsfonts,amssymb,amstext,mathrsfs}
\usepackage{mathpazo}

\usepackage{mathtools}

\usepackage[dvips]{graphicx}
\usepackage{epsf,float}

\usepackage{color,xcolor}
\usepackage{graphicx}
\usepackage{subfigure}
\usepackage{multirow}
\usepackage{tabularx}
\usepackage{pstricks}
\usepackage[section]{placeins}

\usepackage{hyperref}

\usepackage{subfigure}

\newtheorem{property}{Property}

\begin{document}

{\LARGE\centering{\bf{Topological analysis of nuclear pasta phases}}}

\begin{center}
\sf{Rados\l{}aw A. Kycia$^{a,b, }$\footnote{Corresponding author: kycia.radoslaw@gmail.com}, Sebastian Kubis$^{b, }$\footnote{skubis@pk.edu.pl}, W\l{}odzimierz W\'{o}jcik$^{b, }$\footnote{puwojcik@cyf-kr.edu.pl}}
\end{center}

\medskip
\small{
\begin{center}
$^{a~}$Faculty of Science, Masaryk University, \\
Kotl\'{a}\v{r}sk\'{a} 2, 602 00 Brno, Czech Republic. \\
$^{b~}$ Institute of Physics, \\ Cracow University of Technology, \\ Podchor\k{a}\.{z}ych 1, 30-084 Krak\'{o}w, Poland. \\
\end{center}
}
\bigskip

\begin{abstract}
\noindent

In this paper the analysis of the result of numerical simulations of pasta phases using algebraic topology methods is presented. These considerations suggest that some phases can be further split into (sub)phases and therefore should be more refined in numerical simulations. The results presented in the paper can also be used to relate the Euler characteristic from numerical simulations to the geometry of the phases. The Betti numbers are used as they provide finer characterization of the phases. It is also shown that different boundary conditions give different outcomes.

\end{abstract}

PACS 2010: 26.60.-c, 26.60.Gj, 02.40.Re

\section{Introduction}

Numerical modeling of matter under extreme pressure reveals complicated topological phases that were recently presented in \cite{NuclearPastaFormation, DisorderedPasta, NuclearWaffles, Astromaterials} along with many numerical parameters that in principle can provide some information on the nature of phases and their transitions. The next step in understanding such complicated phenomena is performing topological analysis which recovers such results. It can be underlined that this is only the first step in detailed understanding of such phases, however, the analysis presented below, which bases only on topology and hints from numerical simulation can recover characteristic of the material with good accuracy.

\begin{figure}
 \centering
 \includegraphics[width=0.8\textwidth]{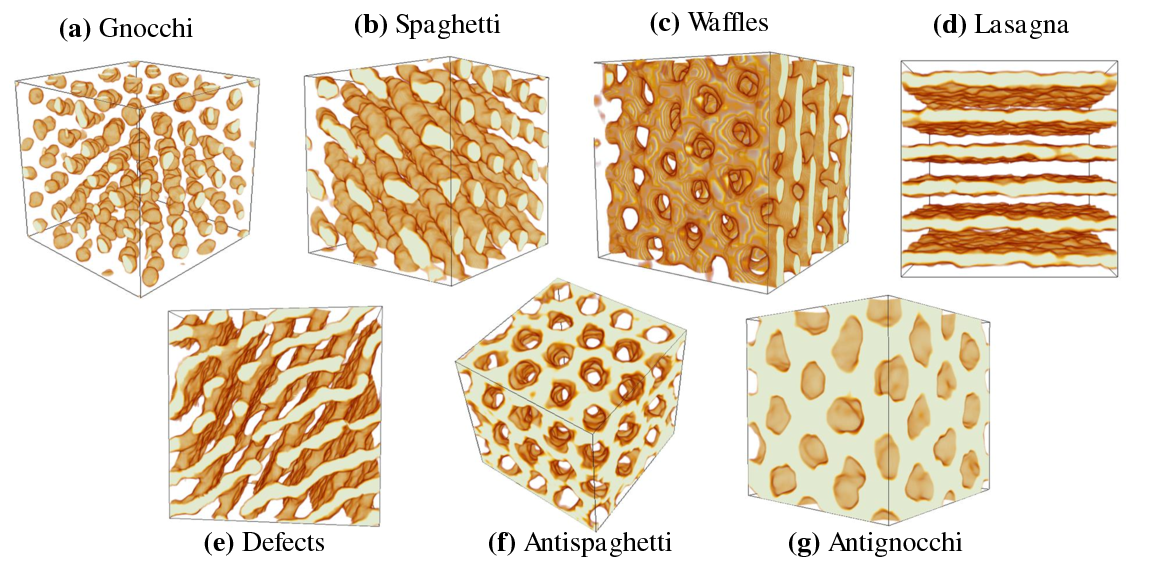}
 \caption{Pasta phases from \cite{Astromaterials}, see also \cite{NuclearPastaFormation, DisorderedPasta, NuclearWaffles}. The plots were reproduced from \cite{Astromaterials} (Fig. 3) with kind permission of Matthew E. Caplan and Charles J. Horowitz.}
 \label{Fig:PastaPhases}
\end{figure}

The numerical results of phases simulations were visualized in \cite{NuclearPastaFormation, DisorderedPasta, NuclearWaffles, Astromaterials} and reprinted in this paper in Fig. \ref{Fig:PastaPhases} for the reader's convenience.
Classification of the phases was there attempted in terms of the Minkowski functionals described in Tab. \ref{Tab:MinkowskiInvariants}.
\begin{table}[tbh!]
\centering
\begin{tabular}{|c|c|}
\hline
 $V$ & volume \\ \hline
 $A=\int_{\partial M}dA$ & surface area \\ \hline
 $B=\frac{1}{4\pi}\int_{\partial M}(k_{1}+k_{2})dA$ & mean breadth \\ \hline
 $\chi = \frac{1}{4\pi}\int_{\partial M} k_{1}k_{2}dA$ & Euler characteristic \\ \hline
\end{tabular}
\caption{The Minkowski functionals from \cite{Astromaterials}. $k_{1}$ and $k_{2}$ are principal curvatures and $K=k_{1}k_{2}$ is the Gaussian curvature and $H=\frac{(k_{1}+k_{2})}{2}$ is the mean curvature.}
\label{Tab:MinkowskiInvariants}
\end{table}
We want to underline that only the Euler characteristic is topological invariant. The other functionals are not invariants, nevertheless the mean curvature - $H$ plays important role in such transitions \cite{SebastianWlodek}.

\begin{figure}
 \centering
 \includegraphics[width=0.8\textwidth]{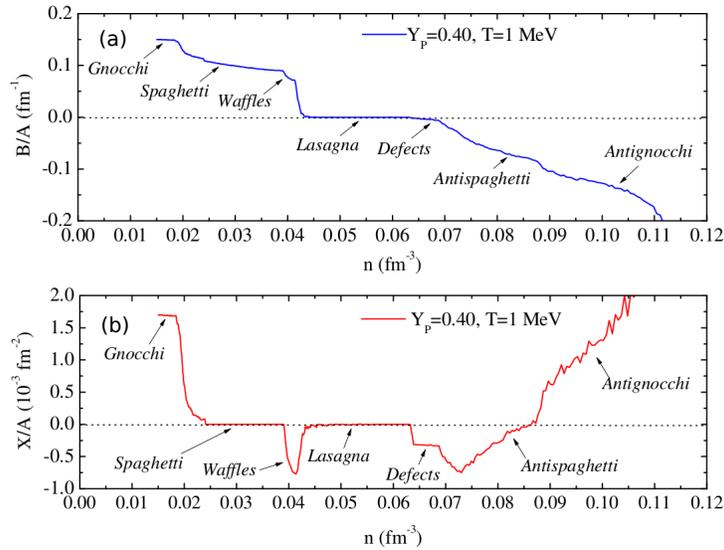}
 \caption{Numerical values of the Minkowski functionals from \cite{Astromaterials}, see also \cite{NuclearPastaFormation, DisorderedPasta, NuclearWaffles}. The parameter $n$ is the density. The plots were reproduced from \cite{Astromaterials} (Fig. 4) with kind permission of Matthew E. Caplan and Charles J. Horowitz.}
 \label{Fig:PastaPhasesPlots}
\end{figure}

The numerical values of the Minkowski functionals are reprinted in Fig. \ref{Fig:PastaPhasesPlots} taken from \cite{Astromaterials}.
Our main result will be to recover and explain the plot of $\frac{\chi}{A}$ from Fig. \ref{Fig:PastaPhasesPlots}. We will focus on the Euler characteristic $\chi$, however division by the surface area normalize it to only one topological component of which the whole structure of the phase is made of. It is useful approach in numerical simulation as it allows to recover (approximately) the value of the quantity (here the Euler characteristic) per one component, under the assumption that the surface of the components do not change drastically. On the other hand the topology cannot recover the surface area as it is theory of invariants under homeomorphisms (continuous mappings with continuous inverse) and the measure (surface area, volume etc.) is not one of them. However, we will see below that general aspects of the plot of  $\frac{\chi}{A}$ is restored in great details.

The most important topological invariant in our considerations will be the Euler characteristic that is the part of algebraic topology. This construction originates from homology theory and detailed introduction can be found in the standard book \cite{Hatcher}. You may also want to consult \cite{Nakahara}, \cite{NashSen} or \cite{Frankel} for more application-oriented exposition.

Our main tool will be the Betti numbers - the $k$th Betti number $b_{k}$ is the rank of the $k$th homology group $H_{k}$ of the surface. As it is well-known \cite{Hatcher} the Betti numbers can be used to define formula for the Euler characteristic in the most general way, namely,
\begin{equation}
 \chi(X) = \sum_{i=1} (-1)^{i} b_{i}=b_{0}-b_{1}+b_{2}-\ldots,
 \label{EulerCharacterisitcBetti}
\end{equation}
where for finite dimensional surface $X$ of dimension $n$ the sum stops at $b_{n}$ as $H_{k>n}=0$, and therefore $b_{k>n}=0$. The Euler characteristic is topology invariant as well as its components - the Betti numbers. However, the Betti numbers are finer characteristic of topological spaces.

The paper is organized as follows: In the next section we describe topological analysis of the models from \cite{Astromaterials} and also \cite{NuclearPastaFormation, DisorderedPasta, NuclearWaffles}. Appendix \ref{Appendix:SpiralDefects} contains description of spiral defect obtained in \cite{DisorderedPasta} and in Appendix \ref{AppendixEmbedding} we discuss the influence of the ambient space (boundary conditions) in which structures are 'embedded' on the Euler characteristic.

\section{Topological analysis}
The simulation cube of \cite{NuclearPastaFormation, DisorderedPasta, NuclearWaffles} has periodic boundary conditions which means that it is topologically a three dimensional torus $T^{3}$. This is an usual choice from technical reasons for simulating infinite space $\mathbb{R}^{3}$ in a finite capacity of computer memory. A brief discussion of the difference between topology of these two approaches is presented in Appendix \ref{AppendixEmbedding}.

In the subsequent subsections we will try to provide the simples and the most intuitive topological analysis for the phases from Fig. \ref{Fig:PastaPhases}. The 'thick' planes will be flatten as topology is unchanged by this operation. We will focus only on one structural element. We will be mostly using invariance of homology groups and therefore the Betti numbers under homotopy (continuous deformation).

\subsection{Gnocchi phase}
First consider the gnocchi phase. There can be two realization. It can be modeled as a sphere $S^{2}$ and then the Betti numbers are $b_{0}=1$, $b_{1}=0$ and $b_{2}=1$, therefore $\chi(S^2) = 2$. When we consider it as a ball $B$, which is contractible to the point, we have $b_{0}=1$ and other Betti numbers vanish. As a result $\chi(B)=1$. From Fig. \ref{Fig:PastaPhases} it seems that the more suitable structure is the ball.

\subsection{Spaghetti phase}
\begin{figure}[htb]
 \centering
 \includegraphics[width=0.8\textwidth]{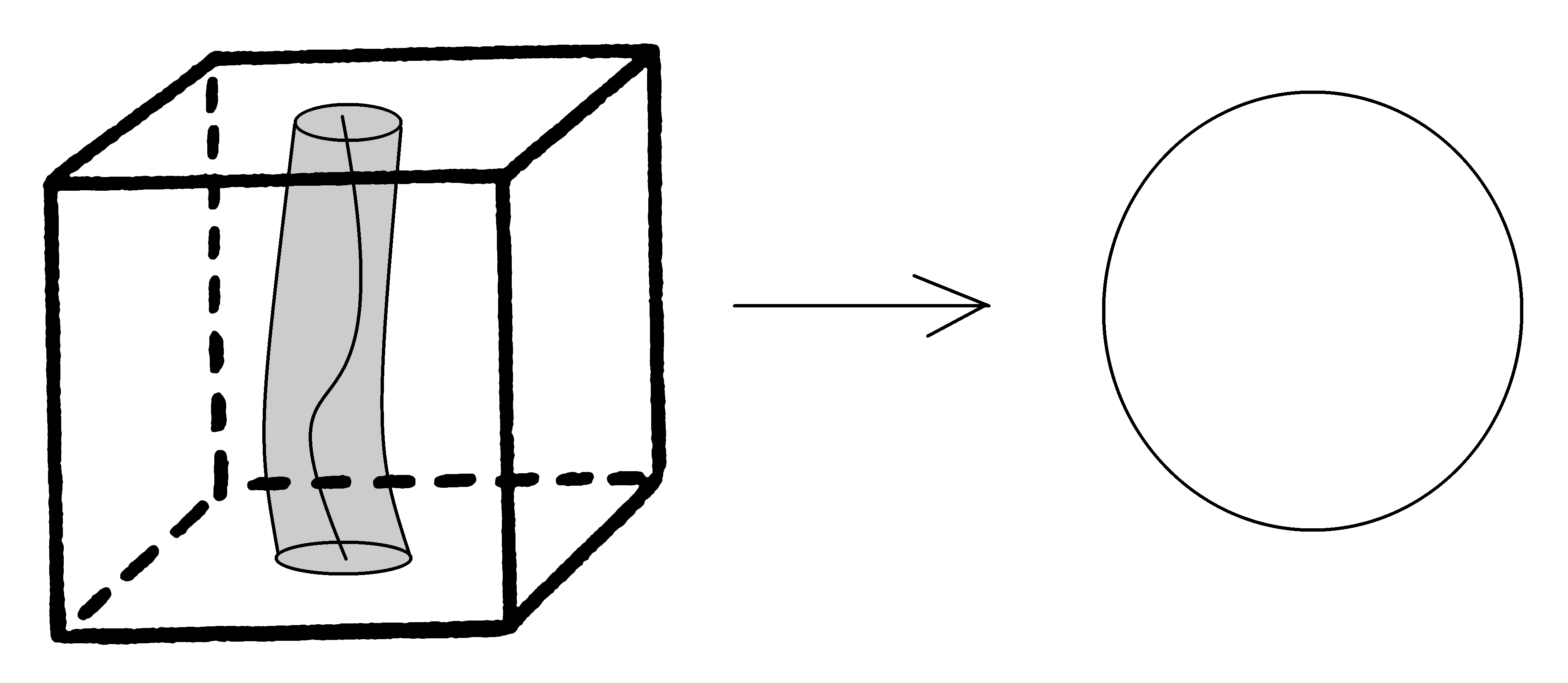}
 \caption{Spaghetti phase topology, which is a circle $S^{1}$.}
 \label{Fig:Spaghetti}
\end{figure}

This phase is represented by the filled tubes attached to opposite sides of simulation cell. Using identification we obtain a filled torus $T^{2}_{f}=S^{1} \times D^{2}$, where $D^{2}$ is the two dimensional disc. We can contract $T^{2}_{f}$ to $S^{1}$ by shrinking the disc $D$ to the point (see Fig. \ref{Fig:Spaghetti}), and the Betti numbers are $b_{0}=1$, $b_{1}=1$; other numbers vanishes. Therefore $\chi(T^{2}_{f})=\chi(S^{1})=0$.

\subsection{Waffles phase and its creation}
Here we present one of the most complicated part of the phase transition as it contains many subphases. It deserves more attention during numerical experiments.

\begin{figure}[htb]
 \centering
 \includegraphics[width=0.8\textwidth]{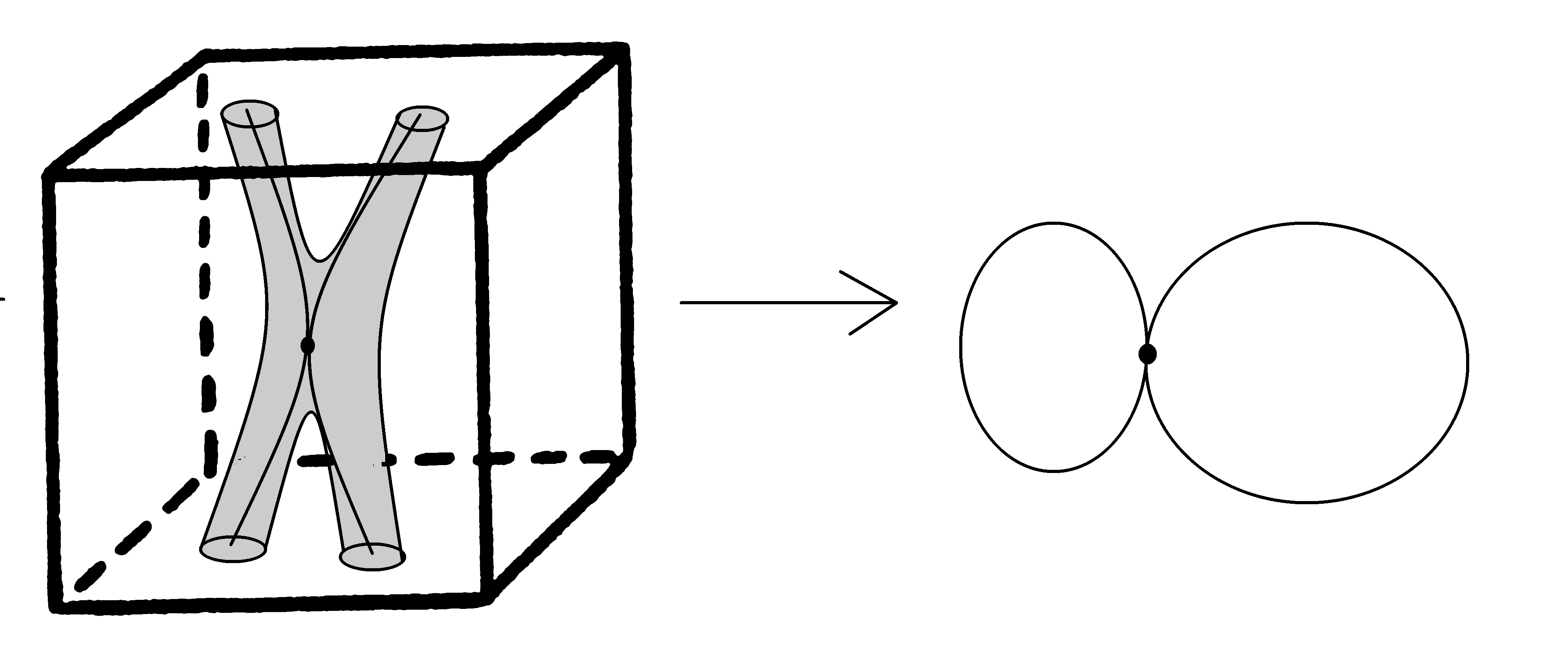}
 \caption{Waffle phase topology - merging of two strings deformed to filled tori $T^{2}_{f} \bigvee T^{2}_{f} \sim S^{1} \bigvee S^{1}$.}
 \label{Fig:Waffle1}
\end{figure}

This structure can be described by shrinking $T^{2}_{f} \sim S^{1}$ that occurred in the Spaghetti phase (see Fig. \ref{Fig:Waffle1}) as such deformation generates the wedge product of the circles - $\bigvee_{i=1}^{n}S^{1}$. From the well-known property, in fact one of the the Eilenberg-Steenrod Axioms \cite{Hatcher}:
\begin{property}
If $X=\vee_{i=1}^{n}X_{i}$ then $H_{k}(X)=\bigoplus_{i=1}^{n}H_{k}(X_{i})$ for $k>0$; $H_{0}(X)$ is the number of connected components of $X$.\footnote{The original axiom is $H_{k}(\coprod_{i}X_{i}) = \bigoplus_{i} H_{k}(X_{i})$ for all $k$, however, we need its form for wedge sum and it requires some modification of $k=0$ case.}
\end{property}
we can easily calculate the homology groups of the wedge product and then the Betti numbers.

That gives $b_{0}=1$, $b_{1}=n$ and zero otherwise, therefore $\chi =1-n$ in this case. 

\begin{figure}[htb]
 \centering
 \includegraphics[width=0.8\textwidth]{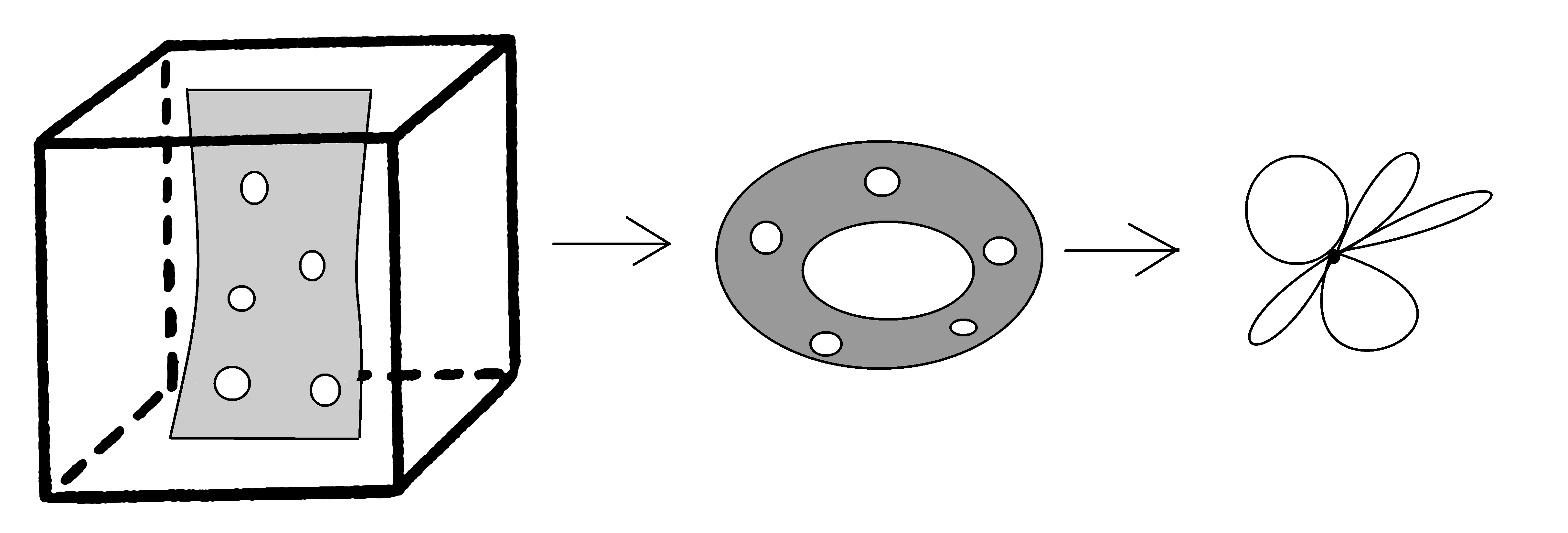}
 \caption{Waffle phase topology - a cylinder (closed 'ribbon') with holes, that can be deformed to $\bigvee_{i=1}^{n}S_{i}^{1}$.}
 \label{Fig:Waffle2}
\end{figure}

At some point it can also be treated as a cylinder with holes - tori can be flatten and holes can be rearranged. This can be further deformed to $\bigvee_{i=1}^{n}S_{i}^{1}$, where $n$ is the number of holes - see Fig. \ref{Fig:Waffle2}. It is the best visible to consider first the cylinder with two holes and expand these holes until they meet each other and then deform them slightly. The Betti numbers then are $b_{0}=1$, $b_{1}=n$ and other numbers vanishes, which produces again $\chi =1-n$.

\begin{figure}[htb]
 \centering
 \includegraphics[width=0.8\textwidth]{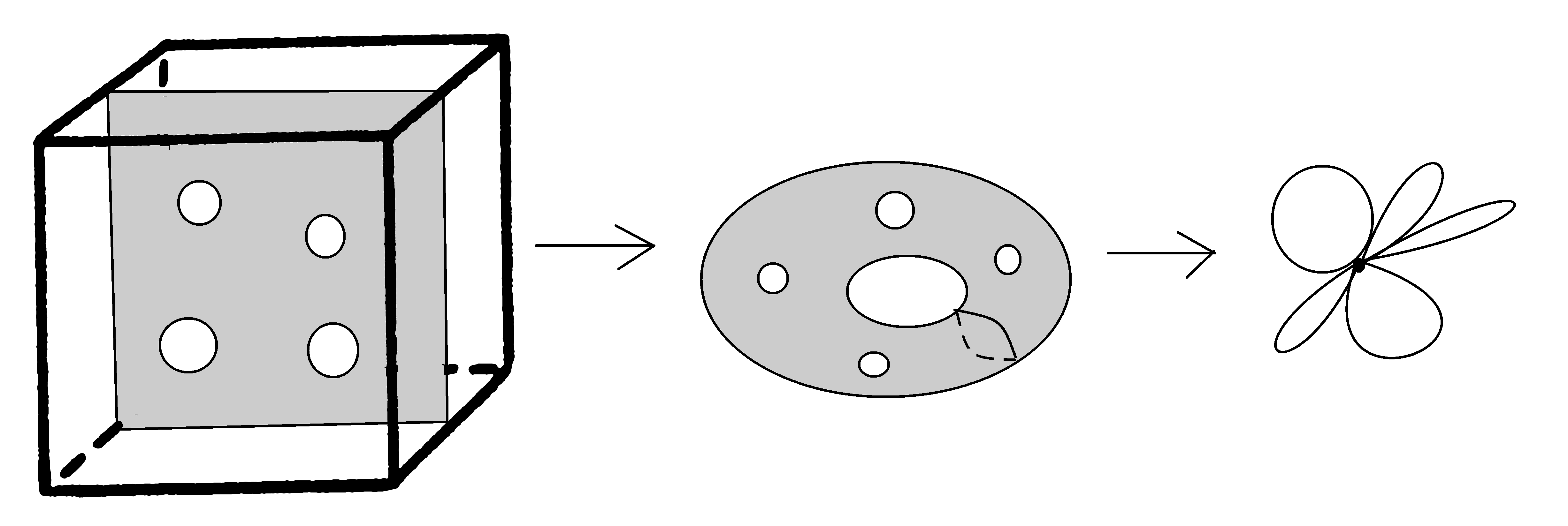}
 \caption{Waffle phase topology - a torus with holes that can be deformed to $\bigvee_{i=1}^{n+1}S_{i}^{1}$.}
 \label{Fig:Waffle3}
\end{figure}

When the multi hole surface is wide enough to reach all four sides of the simulation cell then we can wrap the surface into the surface of the torus with holes. This torus with holes can be deformed into $\bigvee_{i=1}^{n+1}S_{i}^{2}$ for $n$ holes - see Fig. \ref{Fig:Waffle3}. Therefore, again $b_{0}=1$, $b_{1}=n+1$ and other Betti numbers vanishes, that gives $\chi = -n$.

These results suggests that the waffle phase is complicated. The dip in Fig. \ref{Fig:PastaPhasesPlots} marked as waffles can be considered as a complicated multiphase transition and should be examined numerically more carefully.

As each model of 'waffle phase' presented here has Euler characteristic that depends linearly on $-n$ and in numerical simulation the triangle dip can be visible therefore $n$ have to increase when density is increased and then at some point $n$ reverse its direction of change and tends to $0$.

\subsection{Lasagna phase}
\begin{figure}[htb]
 \centering
 \includegraphics[width=0.8\textwidth]{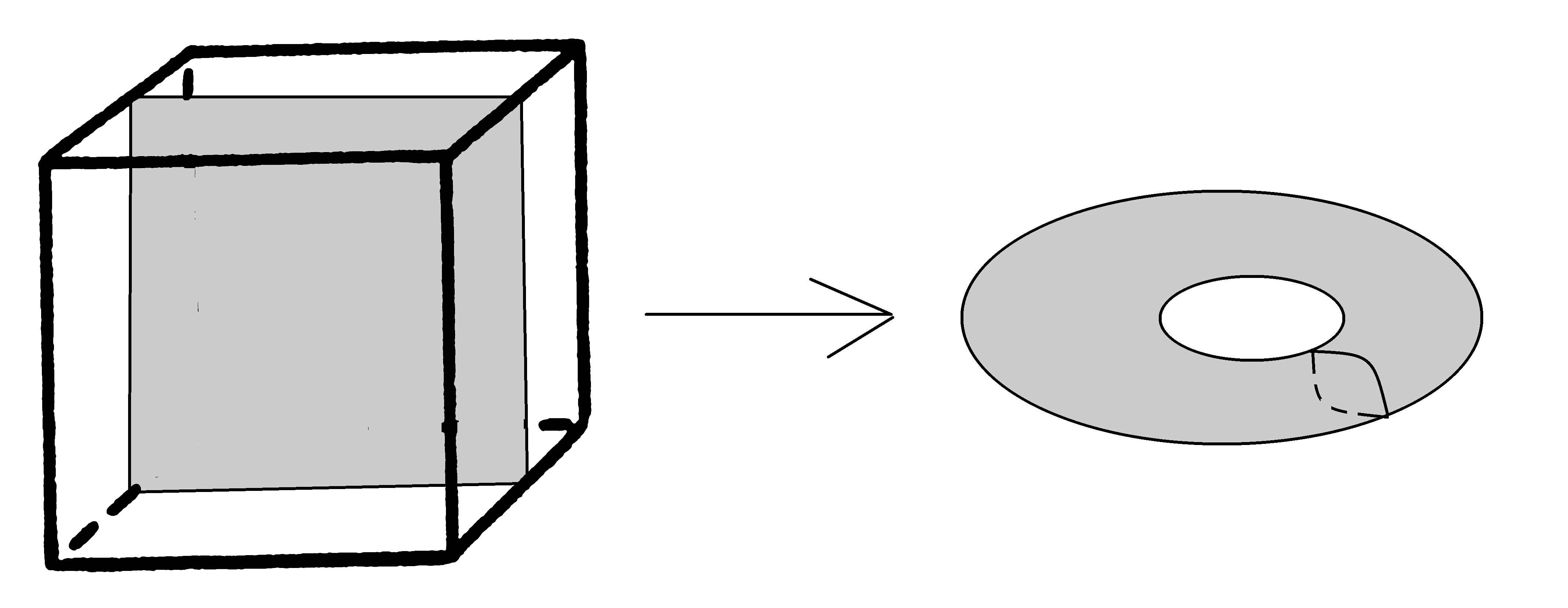}
 \caption{Lasagna phase topology - a torus.}
 \label{Fig:Lasagna}
\end{figure}

When the holes vanishes the parallel planes are obtained, that under identification of opposite simulation cell walls results in tori - see Fig. \ref{Fig:Lasagna}. It is well known fact \cite{Hatcher} that for torus the Betti numbers are $b_{0}=1$, $b_{1}=2$ and $b_{2}=1$ which gives well known result $\chi(T^{2})=2-2g=0$, where the genus $g=1$.

\subsection{Defect phase}
The defect phase is extremely complicated in the view of simple geometrical objects like spheres, planes or tori and nonuniform, therefore, we provide the simplest explanation. Some interesting spiral defect reported in \cite{DisorderedPasta} which is relatively simple to derive topological characteristic, will be presented in Appendix \ref{Appendix:SpiralDefects}.

We assume that the defect appears when some connections between planes appears, which is equivalent to the merging of the tori, which results in increasing the genius $g$ and therefore the Euler characteristics $\chi=2-2g$ decreases. It saturates in Fig. \ref{Fig:PastaPhasesPlots} and therefore additional, non-topological assumption have to be made here.

This model do not describe saturation visible in Fig. \ref{Fig:PastaPhases}, which can indicate some new phase within 'defect phase' set of phases.

\subsection{Antispaghetti phase}
At some point the whole $T^{3}$ simulation cell is filled and situation is reversed - instead of investigating the structures we analyze the topology of holes. Therefore the homology structure of boundary conditions - $T^{3}$ - will be important.

\begin{figure}[htb]
 \centering
 \includegraphics[width=0.8\textwidth]{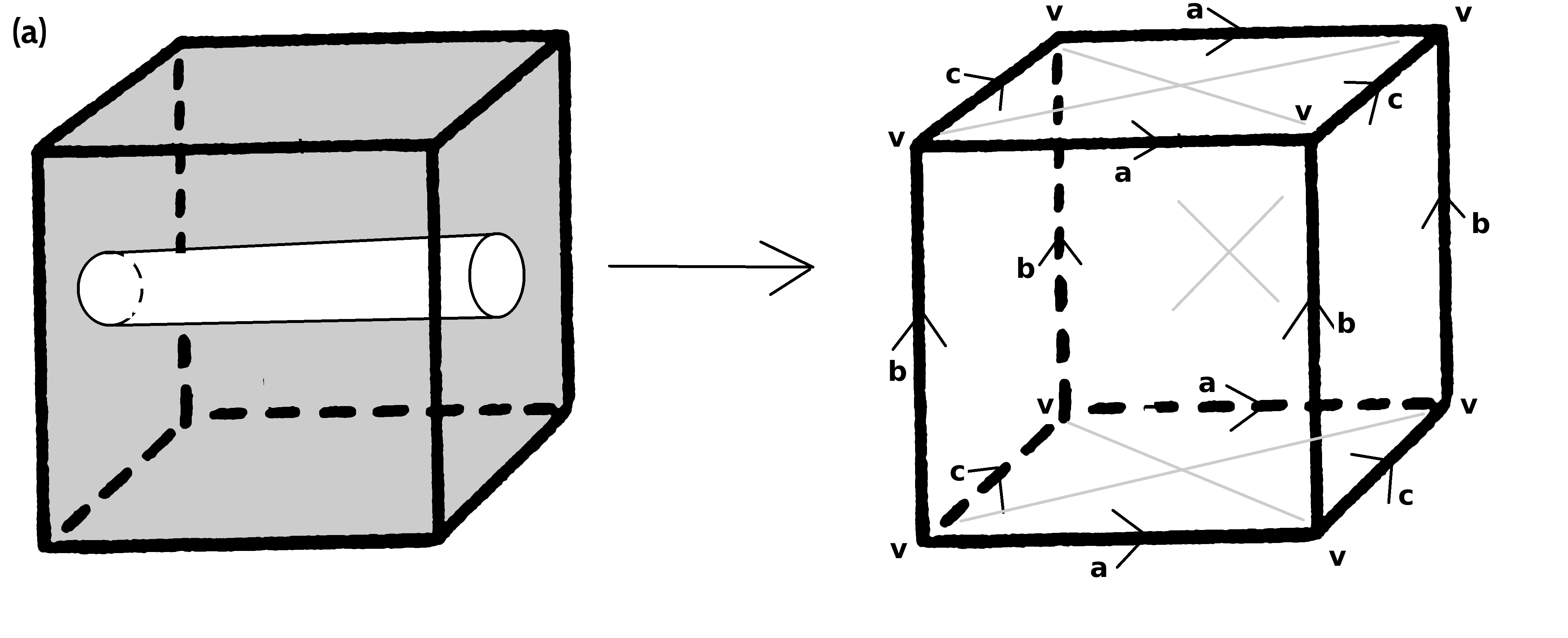}
 \includegraphics[width=0.8\textwidth]{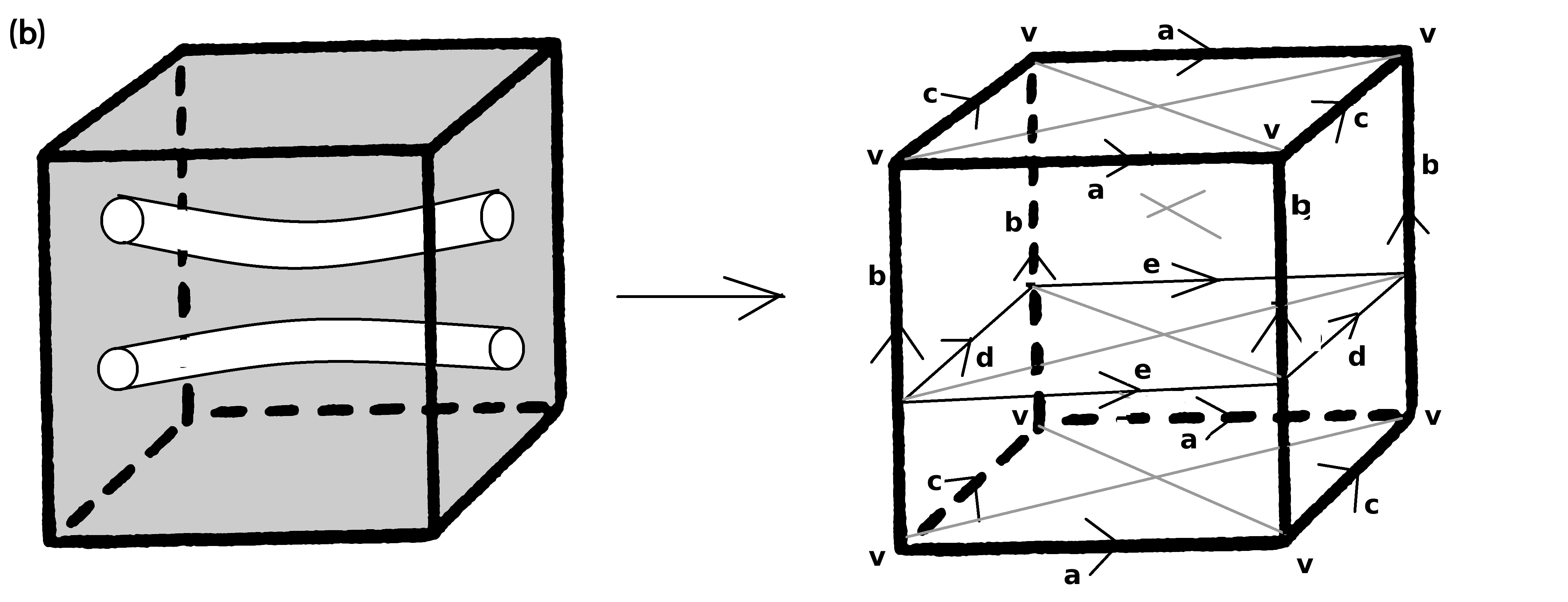}
 \caption{Antispaghetti phase topology - a $T^{3}$ with the holes which are $T^{2}$. For one $T^{2}$-hole by its expansion the right and left face is deformed onto its boundary $[cbc^{-1}b^{-1}]$. For two $T^{2}$-holes  by their expansion left and right faces are deformed onto the torus without 2-cells subdivided by $d$ 1-cell.}
 \label{Fig:Antispaghetti}
\end{figure}

First configuration is the antispaghetti one. Consider the deformation of the torus holes $T^{2}$ cut off from the $T^{3}$ presented in Fig. \ref{Fig:Antispaghetti}.
The algorithm for computing homology of such an object is to expand torus holes until they meet the boundary of the cube and themselves and then use CW-cells structure to compute homology \cite{Hatcher}.

For one $T^{2}$ hole we obtain by deformation, empty cube with identification of opposite faces and the left-right face which is deformed onto its boundary. Therefore the CW-cell structure of homology group has $1 \times$ 0-cell, $3\times$ 1-cell and $2\times$ 2-cell, which gives $b_{0}=1$, $b_{1}=3$ and $b_{2}=2$, and therefore $\chi=0$.

For two $T^{2}$ holes we obtain additional 2-cell inside the cube and only one additional 1-cell $d$ which belongs to the base of homology group, as the $e$ cell is a subdivision of $T^{2}$ which is the cube face glued to it by the word $[aba^{-1}b^{-1}]$. It gives the following multiplicity of homology groups generators $2\times$ 0-cell, $4\times$ 1-cell and $3\times$ 2-cell, that gives $b_{0}=1$, $b_{1}=4$ and $b_{2}=3$, and therefore $\chi = 0$.

Continuing this process inductively, for $n>0$, $T^{2}$-holes we obtain $b_{0}=1$, $b_{1}=3+n$ and $b_{2}=2+n$ that gives $\chi=0$.

\subsection{Antignocchi phase}
\begin{figure}[htb]
 \centering
 \includegraphics[width=0.8\textwidth]{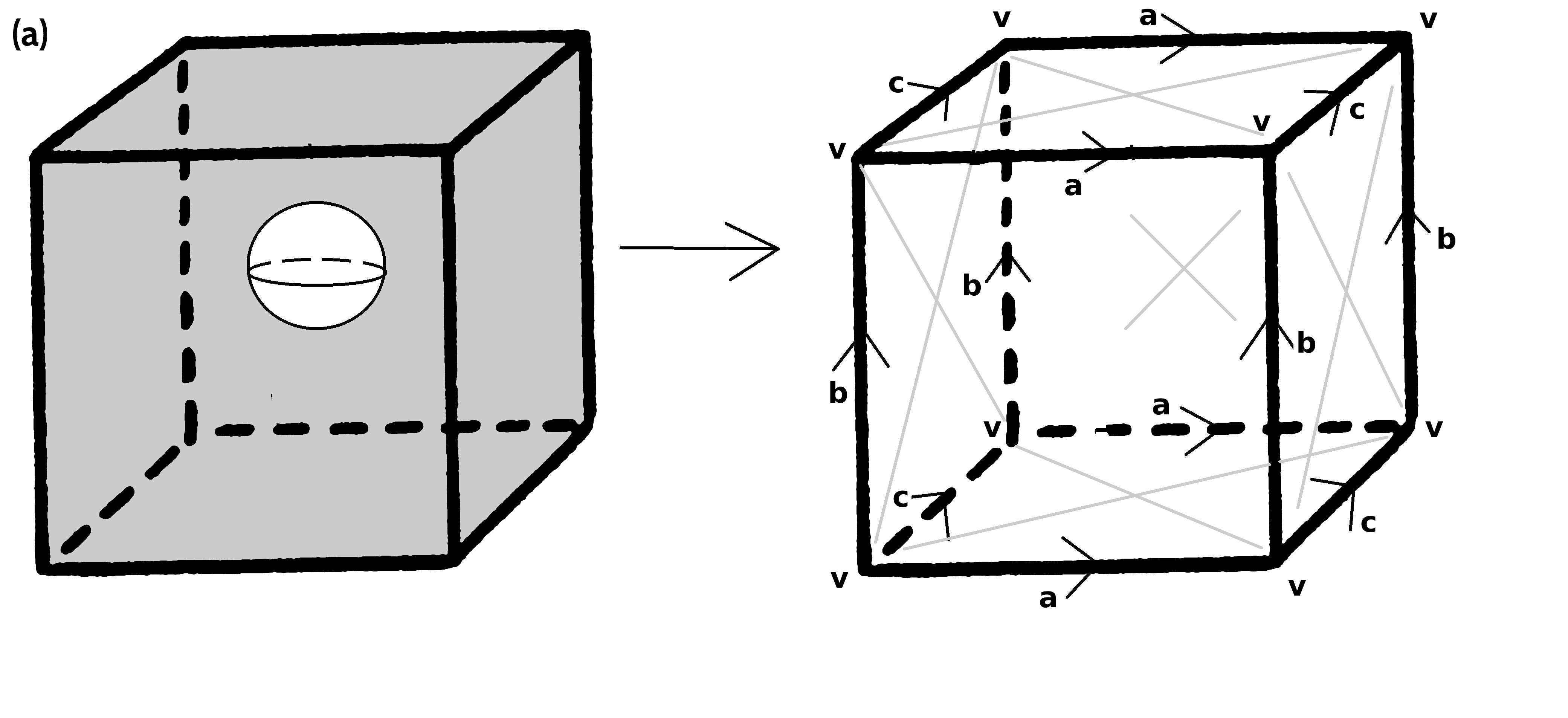}
 \includegraphics[width=0.8\textwidth]{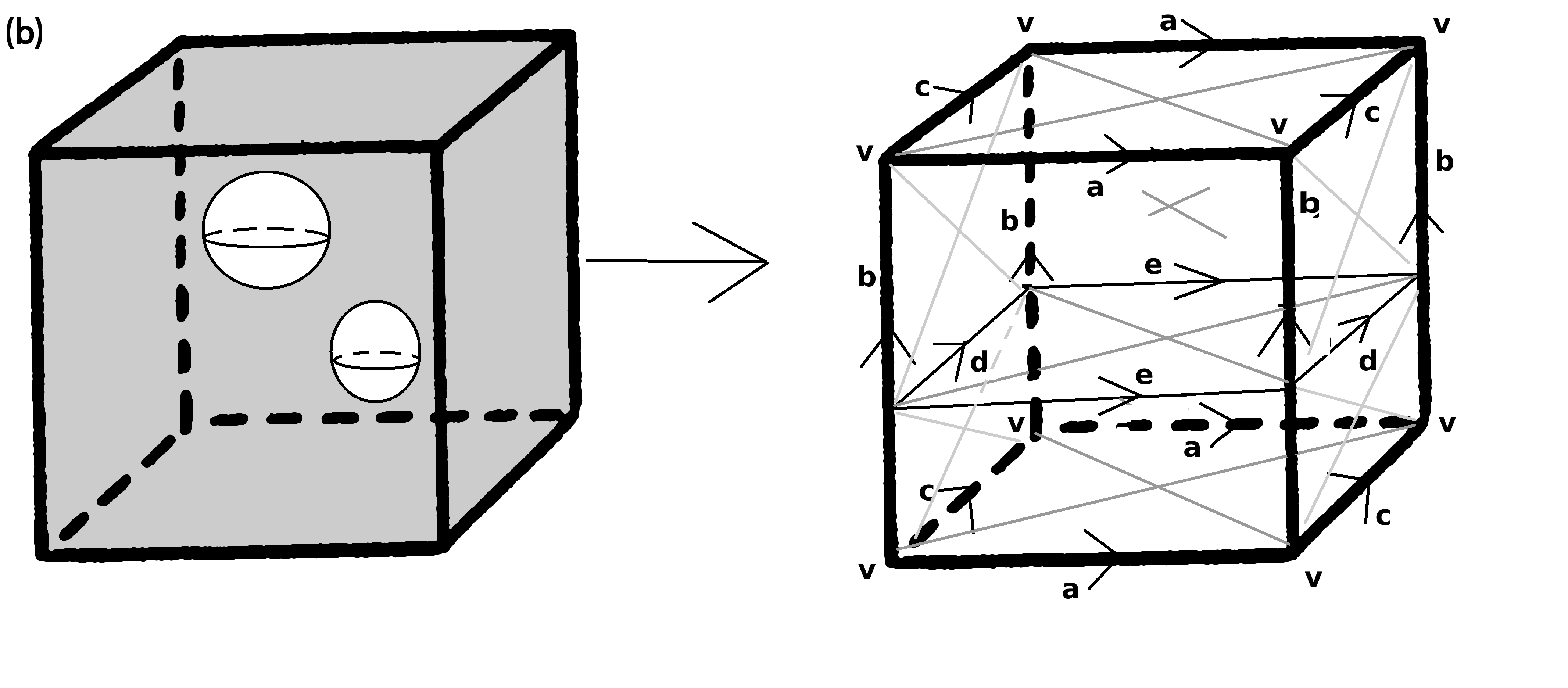}
 \caption{Antignocchi phase topology - a $T^{3}$ with the holes which are two dimensional empty balls.}
 \label{Fig:Antignocchi}
\end{figure}

The final phase consists of $T^{3}$ with holes inside that are empty balls. As for the antispaghetti phase we expand the holes until they meet the boundary and each other, as it is presented in Fig. \ref{Fig:Antignocchi}.
For no holes we obtain $T^{3}$ and $\chi=0$. Further, we can observe that for one hole by expanding the empty ball to the boundary of the cube we obtain $b_{0}=1$, $b_{1}=3$ and $b_{2}=3$ and that gives $\chi = 1$. 

When there are two holes, by expanding them we obtain a cube with identified opposite sides and one 2-cell inside ($T^{2}$) that separates the cube into two parts. The boundaries of this 2-cell - $d$ and $e$ - do not belong to the base of homology group as they form subdivisions of the faces of the cube which are $T^{2}$. Therefore counting the bases of cellular homology groups gives $b_{0}=1$, $b_{1}= 3$ and $b_{2}=4$ and $\chi=2$.

More generally, for $n>0$ spherical holes we obtain $b_{0}=1$, $b_{1}=3$ and $b_{2}=3+n$, and that gives $\chi = 1+n$, a linear growth.

\subsection{Discussion}
\begin{figure}[htb]
 \centering
 \includegraphics[width=0.8\textwidth]{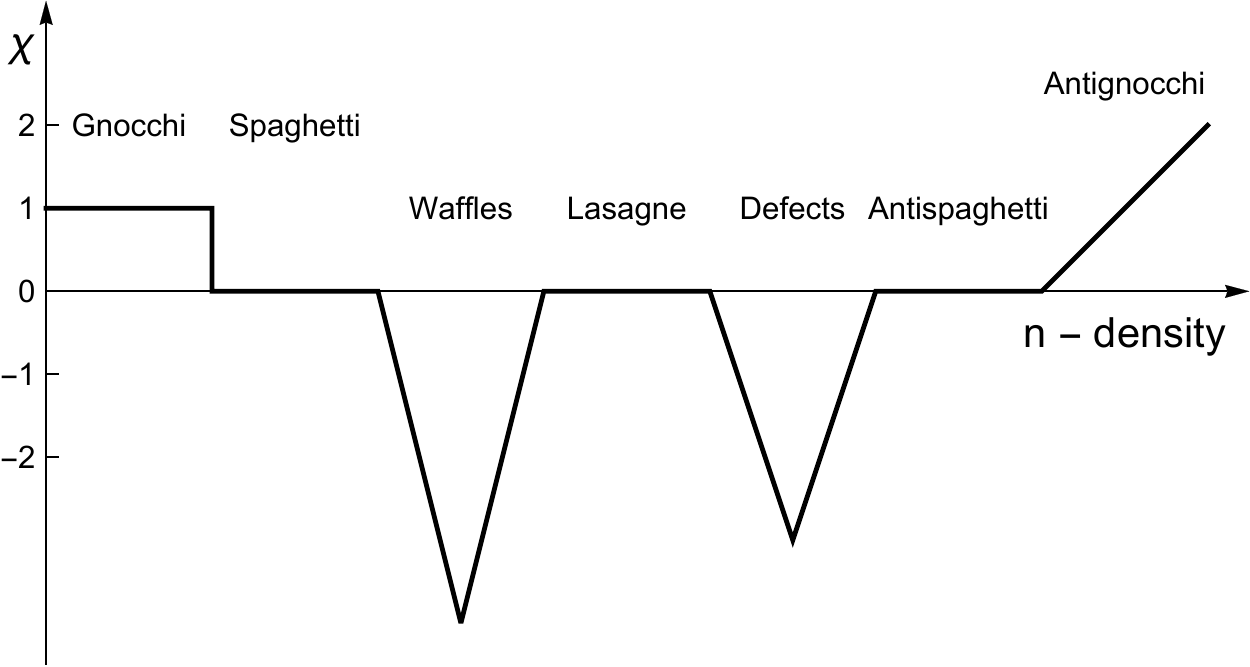}
 \caption{Schematic view of sequence of phases coming from topological analysis. We would like to stress that the values on the plots should be points, as Euler characteristic can change only by integral value, however we plot it as a line to easily compare with numerical results. Here $n$ is the baryon number density as in Fig. \ref{Fig:PastaPhasesPlots}.}
 \label{Fig:ToplogicalPhases}
\end{figure}

In Fig. \ref{Fig:ToplogicalPhases} it is presented the phase distinction based on the above topological analysis.
It seems that the waffle phase consists of a few subphases and should be investigated in more details.

Topological model of the pasta phases gives interesting information about the number of holes in every dimension that are present in the phase. The Euler characteristic combines the number of holes from every dimension and therefore can be considered as more coarse-grained characteristic then the Betti numbers, however all these numbers are topological invariants.

These results shows that knowing the phase and its Euler characteristic we can find out the geometrical characteristics of the phase like the number of holes or the number of planes which are involved in the single structure.

It is also tempting to use the Euler characteristic as a starting point for defining entropy of the system. Such approach was proposed in \cite{TopologicalEntropy} for the $k$-Trigonometric Model(kTM)
\begin{equation}
 S := lim_{N\rightarrow \infty} \frac{1}{N} |\chi|,
\end{equation}
where the the requirement of infinite particles number $N$ in kTM recovers thermodynamical limit. It was used to spot first order phase transitions. From Fig. \ref{Fig:ToplogicalPhases} and \ref{Fig:PastaPhasesPlots} it is highly visible where such a quantity has discontinuity.

\section{Conclusions}
The topological analysis of pasta phases was proposed. It recovers characteristics from numerical simulation without imposing physical contents of the model. Some of phases should be examined better as it seems that additional subphases (in the case of waffles and defects) are present - their description was proposed. 
In antiphases the analysis is not a simple reversal of phases but relays on considerations of boundary conditions. Imposing periodic boundary conditions on the simulation cube makes that gnocchi and antignocchi phases have different Euler characteristics whereas spaghetti and its antiphase have the same. Our topological analysis allows to understand of what is the reason for this behavior obtained in numerical simulations of pasta phases. We want to emphasize that our discussion is independent on the details of the model of interactions.

The topological considerations are good as a starting point for more elaborated discussions involving examination using differential geometry tools as for instance in \cite{SebastianWlodek}.

\section*{Acknowledgement}
We would like to thanks Martin \u{C}adek from Masaryk University, Brno for invaluable suggestions that help to improve the paper. We also would like to thanks Matthew E. Caplan and Charles J. Horowitz for permission to reproduce some of the figures from their papers. Last but not least RK would like to thanks Josef Silhan from Masaryk University, Brno for illuminating discussions on geometry.

\appendix
\section{Appendix}

\subsection{Spiral defects in lasagna}
\label{Appendix:SpiralDefects}
\begin{figure}[htb]
 \centering
 \includegraphics[width=0.8\textwidth]{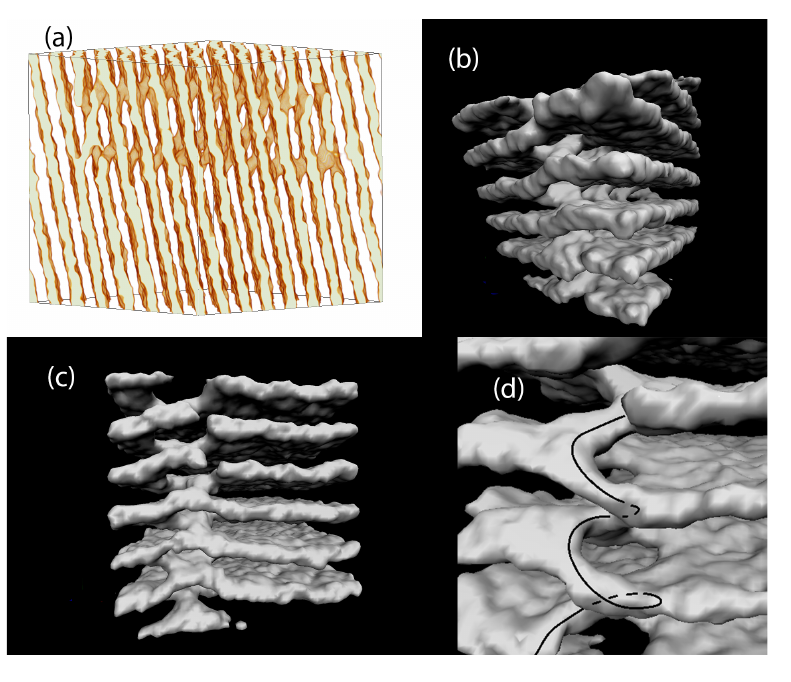}
 \caption{Spiral defect figure taken from \cite{DisorderedPasta}. See especially d). The plots were reproduced from \cite{DisorderedPasta} (Fig. 1) with kind permission of Matthew E. Caplan and Charles J. Horowitz.}
 \label{Fig:SpiralPhasePaper}
\end{figure}

In this short appendix we compute the Betti numbers of spiral defect reported in \cite{DisorderedPasta}, which is reproduced in Fig. \ref{Fig:SpiralPhasePaper}.

\begin{figure}[htb]
 \centering
 \includegraphics[width=0.8\textwidth]{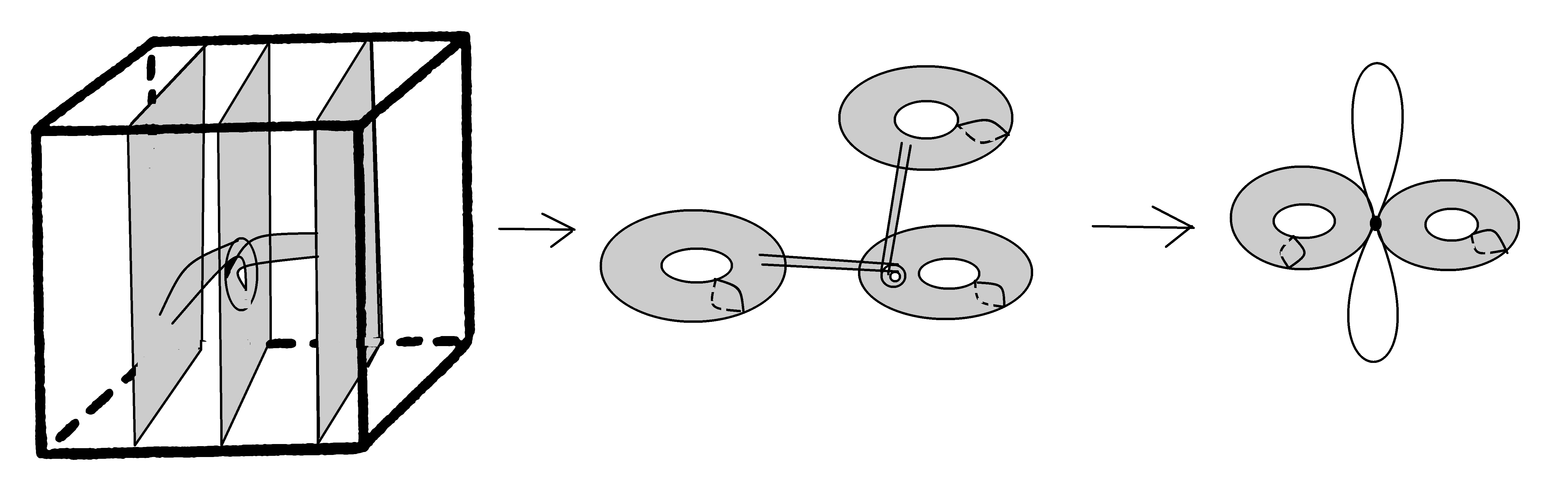}
 \caption{Two boundary planes connected with one middle plane $k=1$ by the spiral ribbon.}
 \label{Fig:Spiral}
\end{figure}

Assume that we cut a holes in $k$ planes (tori) and then glue to that holes a ribbon clockwise and glue the ends to the additional two planes (tori), as presented in Fig. \ref{Fig:Spiral}.
Call the surface $M$, then deforming homeomorphically $k$ tori with holes to the $\bigvee_{i=1}^{2}S^{1}_{k;i}$ and shrinking spiral to the interval and then to the point we get that $M \sim T^{2}\bigvee [\bigvee_{l=1}^{k}(S^{1}_{l} \bigvee S^{1}_{l} )]\bigvee T^{2}$ and adding the Betti numbers of all factors we obtain $b_{0}=1$, $b_{1}=4+2k$ and $b_{2}=2$, therefore, $\chi(M) = -(1+2k) $.

\subsection{Ambient space}
\label{AppendixEmbedding}
In this section we consider influence of 'embedding' of the structures on the Betti numbers. Above the embedding in $T^{3}$ simulation box was discussed, however, the more realistic choice of the ambient space is $\mathbb{R}^{3}$.

It is known that $T^{3}$ is sometimes not well suited for physics. This peculiarity can be seen when considering Maxwell equations on $T^{3}$ - there are solutions with source that results in the electric fields with unbounded value, see \cite{FrankelT3} or \cite{Frankel}, exercise 3.5(2).

We will present results for different choices of ambient space on the example of antispaghetti structure as it is relatively simple to describe.

\subsubsection{$\mathbb{R}^{2}\times S^{1}$}
\begin{figure}[htb]
 \centering
 \includegraphics[width=0.8\textwidth]{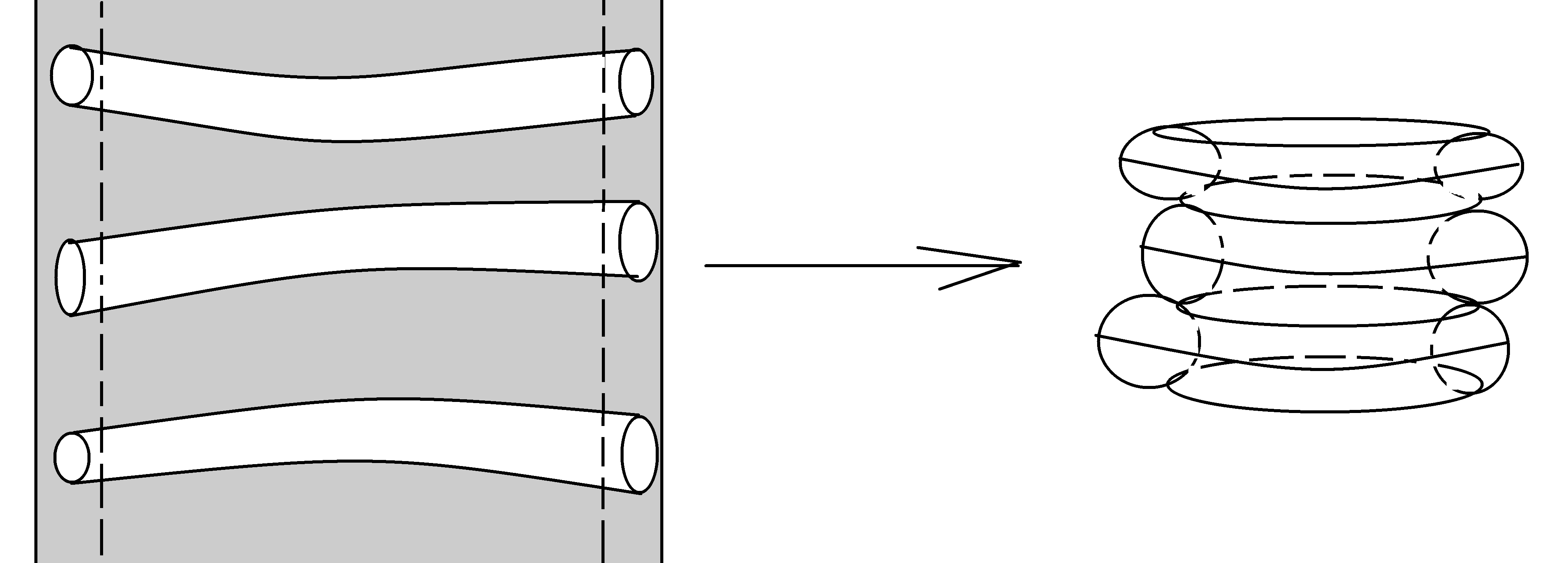}
 \caption{Antispaghetti phase in $\mathbb{R}^{2}\times S^{1}$ ambient space and its deformation for $n=3$ antispaghetti tubes.}
 \label{Fig:AntispaghettiR2S1}
\end{figure}

Consider the $k$ tubes with the surface perpendicular to $\mathbb{R}^{2}$ directions and identified($S^{1}$) along the ends, see Fig. \ref{Fig:AntispaghettiR2S1}.
The structure of the tubes is $(\bigvee_{i=1}^{n}S^{1})\times S^{1}$. The homology groups can be computed using the K\"{u}neth formula for vanishing Tor functor \cite{Hatcher}
\begin{equation}
 H_{0}((\bigvee_{i=1}^{n}S^{1})\times S^{1}) = H_{0}(\bigvee_{i=1}^{n}S^{1})\otimes H_{0}(S^{1})=\mathbb{Z}\otimes \mathbb{Z}\cong \mathbb{Z},
\end{equation}
therefore $b_{0}=1$. Next, 
\begin{equation}
\begin{array}{c}
 H_{1}((\bigvee_{i=1}^{n}S^{1})\times S^{1})= H_{0}(\bigvee_{i=1}^{n}S^{1}) \otimes H_{1}(S^{1}) \oplus H_{1}(\bigvee_{i=1}^{n}S^{1}) \otimes H_{0}(S^{1}) = \\
 =\mathbb{Z}\otimes \mathbb{Z}  \oplus ( \oplus_{i=1}^{n} \mathbb{Z}) \otimes \mathbb{Z} = \oplus_{i=1}^{n+1} (\mathbb{Z}\otimes \mathbb{Z}) \cong  \oplus_{i=1}^{n+1} \mathbb{Z},
\end{array}
\end{equation}
which gives $b_{1}=n+1$. Finally,
\begin{equation}
 H_{2}((\bigvee_{i=1}^{n}S^{1})\times S^{1})=H_{1}(\bigvee_{i=1}^{n}S^{1})\otimes H_{1}(S^{1})=  ( \oplus_{i=1}^{n} \mathbb{Z}) \otimes \mathbb{Z} \cong  \oplus_{i=1}^{n} \mathbb{Z},
\end{equation}
and therefore $b_{2}=n$. As result $\chi = 0$.

\subsubsection{$\mathbb{R}^{3}$}
In the case of Euclidean ambient space we have to consider infinite tubes $S^{1} \times \mathbb{R}$ in $\mathbb{R}^{3}$, there is contraction along all directions from the $n$ infinite tubes to $\bigvee_{i=1}^{n}S^{1}$, which gives $b_{0}=1$, $b_{1}=n$ and other Betti numbers vanishes, and therefore $\chi=1-n$.

These results show that the Euler characteristic as well as Betti numbers strongly depends on the choice of ambient space. Therefore the structure of the simulation boundary conditions can be, in principle, detected by homology invariants.


\end{document}